\documentclass[twocolumn]{pasj00}

\begin{document}

\SetRunningHead{Data-oriented Diagnostics of Pileup Effects on the Suzaku XIS}{S. Yamada et al.}
\title{Data-oriented Diagnostics of Pileup Effects on the Suzaku XIS}

\author{%
Shin'ya \textsc{Yamada}\altaffilmark{1}, 
Hideki \textsc{Uchiyama}\altaffilmark{2}, 
Tadayasu \textsc{Dotani}\altaffilmark{3},
Masahiro \textsc{Tsujimoto}\altaffilmark{3},
Satoru \textsc{Katsuda}\altaffilmark{1}, \\
Kazuo \textsc{Makishima}\altaffilmark{2}, 
Hiromitsu \textsc{Takahashi}\altaffilmark{4}, 
Hirofumi \textsc{Noda}\altaffilmark{2},
Shunsuke \textsc{Torii}\altaffilmark{2}, 
Soki \textsc{Sakurai}\altaffilmark{2}, \\
Teruaki \textsc{Enoto}\altaffilmark{5}, 
Takayuki \textsc{Yuasa}\altaffilmark{3},
Shu \textsc{Koyama}\altaffilmark{6}, 
Aya \textsc{Bamba}\altaffilmark{7}
}

\altaffiltext{1}{
   Cosmic Radiation Laboratory, Institute of Physical and Chemical 
   Research (RIKEN) \\
   2-1 Hirosawa, Wako-shi, Saitama 351-0198}
\altaffiltext{2}{
   Department of Physics, University of Tokyo, 
   7-3-1, Hongo, Bunkyo-ku, Tokyo 113-0033}   
\altaffiltext{3}{Department of High Energy Astrophysics, Institute of 
Space and Astronomical Science (ISAS), \\
Japan Aerospace Exploration Agency (JAXA),
3-1-1 Yoshinodai, Chuo-ku, Sagamihara, Kanagawa 252-5210}   
\altaffiltext{4}{Department of Physics, Hiroshima University,
1-3-1 Kagamiyama, Higashi-Hiroshima, Hiroshima 739-8526}
\altaffiltext{5}{
Kavli Institute for Particle Astrophysics and Cosmology, 
Department of Physics and SLAC National Accelerator Laboratory,\\
Stanford University, Stanford, CA 94305, USA}
\altaffiltext{6}{Department of Physics, Saitama University, 
Shimo-Okubo, Sakura-ku, Saitama-shi, Saitama 338-8570}
\altaffiltext{7}{Department of Science and Engineering, Aoyama Gakuin University, 
5-10-1, Fuchinobe, Chuo-ku, Sagamihara-shi, Kanagawa, 252-5258}

\email{yamada@crab.riken.jp}


\Received{$\langle$Auguest 29, 2011$\rangle$}
\Accepted{$\langle$December 6, 2011$\rangle$}
\Published{$\langle$June, 2012$\rangle$}
\maketitle

\begin{abstract} 

We present the result of a systematic study of pileup phenomena seen in the X-ray Imaging Spectrometer, an X-ray CCD instrument, onboard the Suzaku observatory. 
Using a data set of observed sources in a wide range of brightness and spectral hardness, 
we characterized the pileup fraction, spectral hardening, and grade migration 
as a function of observed count rate in a frame per pixel. 
Using the pileup fraction as a measure of the degree of pileup, 
we found that the relative spectral hardening (the hardness ratio normalized to the intrinsic spectral hardness), 
branching ratio of split events, and that of detached events increase monotonically as the pileup fraction increases, despite the variety of brightness and hardness of the sample sources. 
We derived the pileup fraction as a function of radius used for event extraction. 
Upon practical considerations, we found that events outside of the radius with a pileup fraction of  1\% or 3\% are useful for spectral analysis. 
We present relevant figures, tables, and software for the convenience of users 
who wish to apply our method for their data reduction of piled-up sources.

\end{abstract}

\section{Introduction}
\label{section:1}

Charge coupled devices (CCDs)  have 
been playing an important role in the modern X-ray astronomy (Lumb et al. 1991) 
since the SXT (Soft X-ray Telescope;  \cite{Tsuneta1991}) mounted on YOHKOH.
As a photon-counting detector, 
the first successful use of it in space was realized
with the SIS (Solid-state Imaging Spectrometer; Burke et al. 1991) on board the ASCA satellite. 

The fifth Japanese X-ray satellite Suzaku (\cite{Mitsuda2007}) also carries the CCD devices, 
the X-ray Imaging Spectrometer (XIS; \cite{Koyama2007}), located at the foci of the X-ray Telescope 
(XRT; \cite{Serlemitsos2007}) modules, 
as well as a non-imaging hard X-ray instrument, the Hard X-ray Detector (HXD; \cite{Takahashi2007}). 
The XIS covers soft X-rays from 0.2 to 12 keV, while the HXD covers from 10 to 600 keV. 
The wide-band coverage by a combination of these instruments is a key competence of Suzaku.

Sources that have rich statistics in the HXD are usually bright in the XIS as well. 
In observations of such a bright source, more than one photon can strike the same CCD pixel, 
or one of its immediate neighbors during the exposure time per frame. 
It  causes a complicated and irreversible phenomenon called ``pileup''.  
For example, when two photons with energies of $E_1$ and $E_2$ fall 
in the same pixel within an integration period, 
it is impossible to distinguish them from one with an energy of $E_1 + E_2$. 
It affects obtained spectra, images, and light curves. 
Incorrect evaluation of pileup can lead to wrong interpretations of the data.
This is especially the case for bright sources because their small statistical uncertainty can be overwhelmed 
by systematic uncertainties brought by pileup.
Therefore, it is important to understand pileup effects in order to maximize the wide-band capability of Suzaku. 

In general, when pileup occurs, flux and spectra become lower and harder, respectively. 
The pileup effects of the SISs of ASCA, 
have been studied by \citet{Ebisawa1996}, \citet{Ueda1997}, and Kotani et al. (2000).
Recently, a detailed study for the CCD detector of Chandra, the ACIS, 
as well as a generalized formula of the pileup, 
has been compiled by Davis (2001). 
However, the telescopes used for the XIS and the ACIS are entirely different in the shape and size of the point spread function (PSF), which adds non-linear complications to the effect. 
Therefore, it is difficult to simply apply methods developed for the ACIS to the XIS. 
Likewise, the way of estimating pileup\footnote{http://xmm.esac.esa.int/sas/current/documentation/ \\ threads/epatplot.shtml} for the CCD detector of XMM-Newton, the EPIC, can not be directly applied to the XIS. 
Thus, studies on the pileup effects should be performed specifically for the combined system of XISs and XRTs.

One of the practical solutions to mitigate the pileup in the XIS is the "core-exclusion" method, in which
the central region of the PSF is excluded to avoid the CCD region severely affected by pileup. 
This method was adopted in the analysis of 4U~1630$-$472 (Kubota et al. 2007), Cygnus X-1 (Makishima et al. 2008), GX 339$-$4 (Yamada et al. 2009), and others. 
Based on these experiences, we released a recipe to apply the method
for XIS data 
as an official document\footnote{http://www.astro.isas.ac.jp/suzaku/analysis/xis/pileup/ \\ HowToCheckPileup\_v1.pdf}. 
In the document, it is poorly stated how to estimate the size of the core to be excluded. 
This is often difficult to do by using data that are affected by pileup.

In this paper, we present the results of a systematic analysis for several bright sources affected by pileup in the XIS.
A brief summary on pileup characteristic 
for phenomenological specific to the XIS+XRT system 
is presented in section 2. 
After describing target selection and data processing in section 3, 
systematic studies on pileup effects are presented in section 4.
In section 5, we discuss a criterion that utilizes pileup fraction and 
show concrete examples for the application.  
The main results of the paper are summarized in section 6.

\section{Characteristic of Pileup Effects}

\subsection{Elements that Determine Pileup Effects}

Pileup itself is intrinsically a simple phenomenon. 
However, it is complicated by the following realistic issues: 
\begin{itemize}
\item sizes of charge clouds and pixels, 
\item grade selection both in orbit and on ground,
\item the PSF of the X-ray mirror, 
\item the exposure time per frame, or ``snap time''. 
\end{itemize} 
Therefore, 
the pileup should not be separated from 
the instrumental issues, the X-ray mirror, and 
data-processing specifics. 

When an incident photon is photo-absorbed by a Si atom in the CCD pixel, 
photo electrons
create a charge cloud 
that expands by diffusion along the way to the electrode, 
where the distribution of charges depends on both the absorption depth  and the size of the pixel. 
For any CCD detectors, 
the distributed pattern 
of the charge cloud,``grade'', is used to distinguish between X-ray and non X-ray events.
The pattern changes as pileup becomes more severe.
As easily expected, 
single pixel events are less likely affected by pileup than multiple pixel events.  
If all charge clouds created by a single photon were much smaller than 1 pixel, 
then it would be much easier to reconstruct events. 
However, 
in reality, 
when an  incident photon falls into a border of pixels,
the charge clouds extend larger than 1 pixel, up to four pixels.
Thus, we must consider issues specific to instruments to assess pileup effects. 

\subsection{Features of the XIS + XRT Systems} 

In addition to detailed information on the XIS described in \citet{Koyama2007},  
some features related to pileup effects in the XIS and XRT systems are summarized. 
The XIS system consists of four CCD detectors (XIS0, 1, 2, and 3).
XIS0, 2, and 3 have front-illuminated (FI) CCDs, while XIS1 contains a back-illuminated (BI) CCD.
The depletion layers of the FI and BI CCD are $65~\mu $m and $42~ \mu $m, respectively. 
XIS2 has been dysfunctional since November 2006. 
Among the remaining CCDs, XIS3 is almost identical to XIS0, while XIS1 is different. 
Therefore, we treat XIS0(3) and XIS1 differently throughout this paper, unless otherwise noted. 

Events with grade 0, 2, 3, 4, and 6 are regarded as X-ray events, 
while events with other grades as cosmic-ray events (see Koyama et al. 2007 and figure 4 of Yamaguchi et al. 2006). 
The grade 1 and 5 events are charge distribution called ``detached events''. 
As pileup becomes more severe, 
more X-ray events are regarded as grade 1, 5, and 7, 
because of the overlaid distributions of multiple charge clouds.
Note that, in the case of the normal clocking mode, the XIS has three editing modes, 2$\times$2, 3$\times$3 and 5$\times$5, 
which means that the pulse height values of the number of pixels are sent as a telemetry. 
The event detection algorithm is, however, common to all the
editing modes; the central pixel and the surrounding 8 pixels
are used to detect an event on orbit. 
On the other hand, 
the grading algorithm utilizes not only the 3x3 pixels of the
event but also outer pixels, whose details depend on the
editing modes.
Note that most X-ray events, which are photo-absorbed not in neutral region but in a depletion layer, 
are not split into a region larger than 2$\times$2 pixels.

The pileup effects primarily depend on one parameter, 
the number of counts per snap time in a detection cell. 
In the case of the XIS, 
the number of detection cells is nine (3$\times$3);
i.e., 
the central pixel and the surrounding 8 pixels are used to detect the event in orbit. 
Efforts to mitigate pileup in the XIS are thus divided largely into two: 
(1) to reduce the effective frame time (snap time) by partial read of the CCD 
and (2) to use events in the outskirts of the PSF, 
in which the surface brightness is lower than that at the core. 

Each XIS has two clocking modes: P-sum and Normal.   
Throughout this paper, we treat data taken only with the latter mode. 
Snap time is basically 8 s in the Normal clocking mode. 
The XIS has two options that make the snap time shorter than 8 s to reduce counts per a frame time.
One of them is the window option, in which  
only a certain part of the CCD, 
called window, is used for readout. 
The other is the burst option, 
in which the snap time can 
be shortened arbitrarily by discarding events arriving in a certain period during a frame time.
An interval between the end of the snap time and the start of the next one is called ``delayed  time''.  
The delayed time is 0 s when the burst option is off, 
while it becomes longer than 0 s when the burst option is employed.
The window and burst options 
can be applied independently for each sensor.

The combination of the XIS and the XRT provides 
a spatial resolution of $\sim\timeform{2\prime}$ in a half-power diameter (HPD). 
Note that one pixel of the XIS corresponds to $1''.042$. 
The field-of-view (FOV) is $\timeform{18\prime} \times \timeform{18\prime}$  
at the maximum, which becomes smaller when a window option is used;  
$\timeform{18\prime} \times (1/4\times\timeform{18\prime})$  for 1/4 window option, 
and $\timeform{18\prime} \times (1/8\times\timeform{18\prime} ) $  for 1/8 window option. 
Due to the attitude fluctuation and calibration uncertainties, 
the 1/4 window is recommended over the 1/8 window option.
In many cases, a combination of the 1/4 window and burst options is used to mitigate the pileup. 
A rule of thumb for the count rate to cause a possible pileup for point sources 
for the Normal full window mode is $\gtrsim 12$ s$^{-1}$ 
(see 7.9.1 of Suzaku technical description\footnote{http://www.astro.isas.ac.jp/suzaku/doc/suzaku\_td/\\node10.html\#SECTION001091000000000000000}). 

The XIS+XRT system 
has two features that make the core-extraction method practically viable; 
(1) energy-independence of the PSF shape \citep{Serlemitsos2007}, and (2) heavy over-sampling.
In other words, we can perform experiments of changing the surface brightness 
just by employing the different extraction radius from the PSF core, 
keeping the other instrumental responses unchanged. 
The CCD detectors onboard Chandra and XMM-Newton lack either or both of the two features, 
which makes the application and validation of the method more complicated.

\subsection{Observed Features Caused by Pileup Effects}

As a result of convolving mirror and detector responses with piled-up events, 
the following observational features appear. 

\vspace{0.15cm}
\noindent {\bf Image} \\
\vspace{0.07cm}
The observed peak of the PSF is smeared out. 
The smearing is often difficult to distinguish from the PSF core 
flattening due to the thermal wobbling (Uchiyama et al. 2008). 
In some cases,
a hole at the center is created 
by the two reasons: 
(1) an event is regarded 
as an non X-ray grade due to many events falling around it, 
and (2) when, 
pileup becomes more severe, 
a sum of pulse height of piled-up events exceeds an 
upper threshold used to determine an event, and it is not regarded as an event.

\vspace{0.1cm}
\noindent {\bf Spectra}\\
\vspace{0.07cm}
The measured count rate decreases in lower energy ranges, 
while it increases in the higher energy ranges, 
which artificially creates a hard tail in the spectrum. 
Since this feature is convolved with the intrinsic spectra, 
it appears more clearly in a source with softer spectra than one with harder spectra, 

\vspace{0.1cm} 
\noindent {\bf Light curves}\\ 
\vspace{0.1cm}
When the flux of the source changes during an observation, 
the light curve can become plateaued at the peak. 

\vspace{0.1cm}
\noindent {\bf Grade}\\ 
\vspace{0.07cm}
The branching ratio among the grades can change. 
In general, 
pileup decreases single-pixel events (grade 0) and increase multiple-pixel events (grade 2, 3, 4, 6, and 7). 

The extents of these phenomena depend on incident flux and spectra of a source. 
Based on these observational clues,  
we construct reliable criteria to judge and exclude the pileup effects.

\subsection{Definition of the Pileup Fraction} 

In order to characterize the degree of pileup, 
we use the pileup fraction $f_{\mathrm{pl}}$ following Davis et al. (2001).
The definition of $f_{\rm{pl}}$ is 
the ratio of $P(k \ge 2, x)$ to $P( k \ge 1, x)$, where $P(k,x)$ is the Poisson distribution function, 
$x$ is defined as the mean counts per snap time per detection cell, 
and $k$ is the actual number of photons 
falling per snap time per detection cell. 
It is expressed as 
\begin{equation}
\displaystyle f_{\rm{pl}}(x) = \frac{ \Sigma_{k=2}^{\infty} P(k,x) }{ \Sigma_{k=1}^{\infty} P(k,x)} = \frac{1}{2} x - \frac{1}{12}x^2 + O(x^4).
\end{equation}
We expanded the equation by the second order of $x$ since $x$ is in many cases smaller than one. 

The theoretical value of $f^{\rm{th}}_{\rm{pl}}$ can be derived 
by substituting $x$ with the incident count. 
However, when pileup occurs, 
the incident count is unknown from the data. 
Therefore, we use the observed piled-up counts ($x_{obs}$) 
and an observed pileup fraction $f^{\rm{obs}}_{\rm{pl}}$ as a proxy for a theoretical one $f^{\rm{th}}_{\rm{pl}}$.

\section{Data Set and Processing}
\subsection{Target Selection}

To perform a systematic study on pileup, 
we need to select as many point sources as possible to cover wide ranges of 
brightness of photons and intrinsic hardness, 
because the extent of pileup is determined by the former, 
while the appearance of pileup effects is sensitive to the latter. 

The list of the targets, 
as well as observational parameters of these sources (e.g., the snap and delayed times)
are summarized in table \ref{sample}.
These meet the following criteria: 
(1) brighter than $\sim$ 1 mCrab and (2) small variability with a root mean square of the count rate smaller than 30\% of its mean. 
The latter criterion is needed to avoid highly varying sources, 
in which pileup effects are difficult to disentangle with varying flux and spectra.

\begin{table*}
\caption{The basic information and the operated modes of the XISs for the selected samples.}
 \label{sample}
 \begin{center}
  \begin{tabular}{lccrrrrrrr}
  \hline\hline
\multicolumn{1}{c}{Name$^\S$} & Obsid & Date & $T$ (ks)$^*$ & \multicolumn{2}{c}{XIS0} & \multicolumn{2}{c}{XIS1}  & \multicolumn{2}{c}{XIS 3}  \\[2mm] 
   &             &          & & S/D$^\dagger$ (s)  & \multicolumn{1}{c}{Rate}$^\ddagger$ & S/D (s) & \multicolumn{1}{c}{Rate} & S/D (s) & \multicolumn{1}{c}{Rate} \\[2mm]                      
\hline
GX~17+2 & 402050020 & 2007-09-27 & 56.3 & 0.5/1.5 & 327.5 & 2.0/6.0 & 346.6 & 0.5/1.5 & 375.5 \\
Serpens~X-1 & 401048010 & 2006-10-24 & 81.3 & 1.0/1.0 & 259.0 & 1.0/1.0 & 281.4 & 1.0/1.0 & 263.1 \\
GX~339$-$4 & 401068010 & 2007-02-12 & 256.5 & 0.3/1.7 & 950.5 & 0.3/1.7 & 936.3 & 0.5/1.5 & 866.7 \\
Crab & 101004020 & 2006-04-04 & 40.4 & 0.1/1.9 & 1739.6 & 0.1/7.9 & 2801.6 & 0.1/1.9 & 1710.6 \\
Cygnus~X-1(\#1) & 404075120 & 2009-06-04 & 44.5 & 0.5/1.5 & 412.8 & 0.5/1.5 & 500.8 & 0.5/1.5 & 472.4 \\
4U~1705$-$44 & 401046020 & 2006-09-18 & 29.8 & 2.0/0.0 & 88.4 & 2.0/0.0 & 86.2 & 2.0/0.0 & 91.0 \\
GX~349+2 & 400003010 & 2006-03-14 & 58.4 & 0.3/0.7 & 339.9 & 0.3/1.7 & 498.9 & 0.3/0.7 & 401.5 \\
Cygnus~X-1(\#2) & 404075130 & 2009-10-21 & 40.0 & 0.5/1.5 & 228.1 & 0.5/1.5 & 267.9 & 0.5/1.5 & 262.2 \\
4U~1630$-$472 & 400010010 & 2006-02-08 & 43.8 & 1.0/1.0 & 112.0 & 1.0/1.0 & 116.2 & 1.0/1.0 & 127.8 \\
Aql~X-1 & 402053010 & 2007-09-28 & 35.7 & 0.5/1.5 & 159.6 & 0.5/1.5 & 187.2 & 0.5/1.5 & 162.9 \\
Vela~X-1 & 403045010 & 2008-06-17 & 147.4 & 2.0/0.0 & 39.0 & 2.0/0.0 & 38.9 & 2.0/0.0 & 40.0 \\
3C~273 & 702070010 & 2007-06-30 & 109.2 & 8.0/0.0 & 5.1 & 8.0/0.0 & 7.6 & 8.0/0.0 & 5.5 \\
4U~1636$-$536 & 401050010 & 2007-02-09 & 52.1 & 1.0/1.0 & 49.2 & 1.0/1.0 & 56.7 & 1.0/1.0 & 48.7 \\
MCG~--6-30-15 & 700007010 & 2006-01-09 & 359.9 & 8.0/0.0 & 3.9 & 8.0/0.0 & 6.6 & 8.0/0.0 & 3.7 \\
SS~Cyg & 400006010 & 2005-11-02 & 80.8 & 8.0/0.0 & 3.8 & 8.0/0.0 & 6.4 & 8.0/0.0 & 3.7 \\
MCG~--5-23-16 & 700002010 & 2005-12-07 & 219.6 & 8.0/0.0 & 3.3 & 8.0/0.0 & 4.8 & 8.0/0.0 & 3.8 \\
PKS~2155$-$304 & 104004010 & 2009-05-27 & 157.9 & 2.0/0.0 & 4.8 & 2.0/0.0 & 9.7 & 2.0/0.0 & 5.1 \\
\hline\hline
\end{tabular}
\end{center}
\begin{itemize}
\item[$^*$] Time from start to end of the observation.
\item[$^\dagger$]  `S and `D' refer to snap time and delayed time in units of seconds, respectively.  
\item[$^\ddagger$]  The count rate integrating over a whole region per seconds. 
\item[$^\S$] Names are sorted by ascending order of counts per snap time in the XIS1.
\end{itemize}
\end{table*}

\subsection{Data Processing and Event Screening}

We performed data processing from unscreened event files, 
and first discarded the telemetry-saturated periods by using the house keeping data\footnote{
The file {\tt aeNNNNNNNNNxi[0$-$3]\_0\_tel\_uf.gti} can be found in the distributed processing products.
} provided by the XIS team. 
The telemetry saturation itself does not change the spectral shape,   
though it may lead to underestimation 
of flux, and hence causing wrong estimation of pileup in some cases. 

We further screened the events in the same way as the standard criteria 
based on e.g., conditions on the satellite orbit and data rates\footnote{See http://www.astro.isas.jaxa.jp/suzaku/process/\\v2changes/criteria\_xis.html.}. 
For the analysis of grade-branching ratio, 
we have extracted grade 02346, grade 0, grade 2346, and grade 1, and grade 5 events, separately. 

\subsubsection{Attitude correction}

\begin{figure}[htbp]
    \begin{center}
      \includegraphics[width=0.48\textwidth]{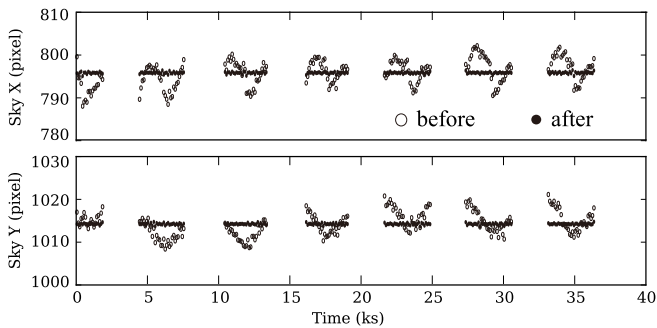}	
            \end{center}
    \caption{Time profiles of the image center in the SKY X and Y coordinate in the unit of pixel, taken from observation of Cygnus X-1 performed on October 21 in 2009. 
    The bin size is 100 s. The open and filled circles are the values of SKY Y and X before and after applying the attitude correction with {\tt aeatttuned.py}, respectively.}
    \label{attcor}
\end{figure}

As mentioned in Uchiyama et al. (2008), 
the attitude of the Suzaku satellite can be restored by 
using house keeping data including the temperature 
which has been implemented in the \texttt{aeattcoord} tool. 
Figure \ref{attcor} shows the time profiles (open circles) of the barycenter position of Cygnus X-1 (\#2) calculated from the observed image
 in the SKY X and Y  (the coordinate corresponding to the position on the celestial sphere). 
Although the attitude was corrected with {\tt aeattcoord} to some extent, it still shows a fluctuation of $\sim$ 20 pixel in a time scale of $\sim$ 1 ks. 
This fluctuation is non-negligible for evaluating the pileup effects at the peak of the PSF.

We further corrected the image for the residual fluctuation using 
our python script \texttt{aeatttuned.py}, 
which has the same function as {\tt aeattcor2}\footnote{http://heasarc.gsfc.nasa.gov/ftools/caldb/help/aeattcor2.html} in the HEAsoft package
provided by John E. Davis, M. Nowak, and others in MIT. 
This method (1) creates the time profile of the image center of SKY X and Y in a time bin of 100 s, 
(2) modifies the attitude file to correct for the fluctuation.
and (3) reprocesses the event file with the modified attitude file using {\tt xiscoord}. 
We show the time profiles (filled circles) in figure \ref{attcor} obtained after applying this method.
Through this process, the fluctuation of SKY X and Y 
decreases to $\sim$ 3 pixel. 
In the following analysis, 
we have applied the attitude correction for all the events unless otherwise noted. 
Note that the method functions properly 
only when a count rate is high enough to track image centers for every 100 s.

\subsubsection{Background subtraction}

We used the observation of the North Ecliptic Pole (NEP) 
to estimate the background count rate, taken on October 16 in 2007 (OBSID 100018010), 
considering the difference in a snap time. 
The counts per snap time (8 s in the case of NEP) per pixel of XIS0 and XIS1 
are $1.6 \times 10^{-6}$ and $3.8 \times10^{-6}$ counts frame$^{-1}$ pixel$^{-1}$, respectively. 
This is much smaller than most sources in our targets, 
but can be non-negligible of $\sim$10\% of the surface brightness at the PSF outskirt 
of the dimmest sources in our list.

Since the NXB of the XIS changes by a factor of $\sim$ 2 (Tawa et al. 2008), 
the systematic error in the subtraction of the NXB is expected to be $\sim$20\% at most. 
In the following analysis, 
we have estimated backgrounds from the NEP for 
corresponding grade selections, otherwise stated. 
When creating a radial profile of the surface brightness, 
we subtracted values in the NEP from those of the sources in each radial bin.

The Cosmic X-ray Background (CXB) was not subtracted from the spectra, 
since it is almost negligible of less than 0.1\% of all the source signals in any energy ranges. 
Hot pixels are excluded from all of the events throughout the paper.  

\begin{figure*}[htbp]
    \begin{center}
      \includegraphics[width=0.95\textwidth]{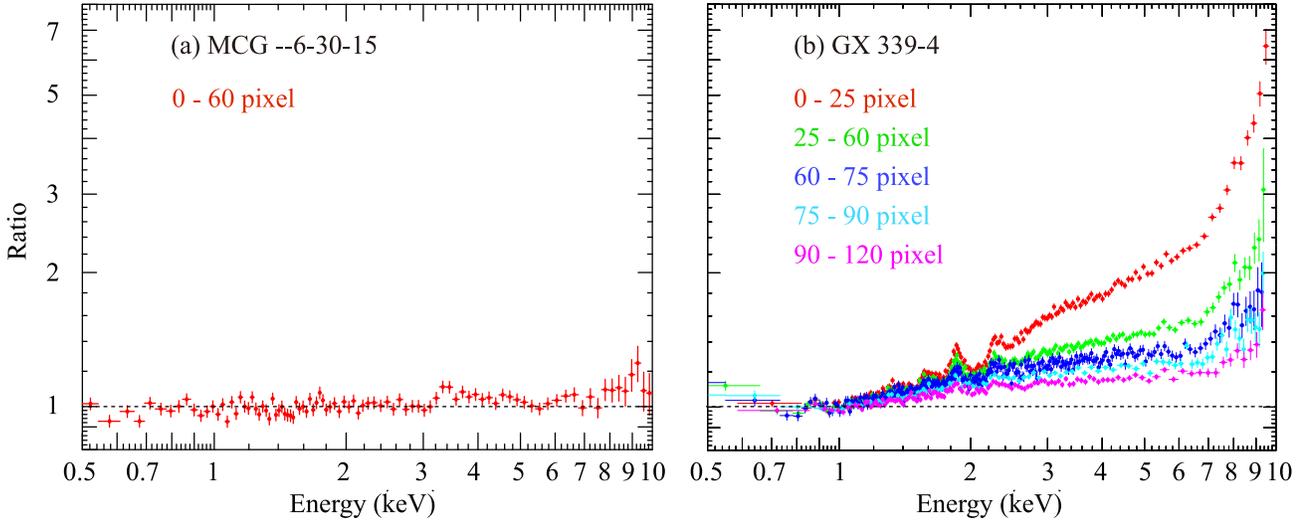}	
            \end{center}
    \caption{    
    Ratio of XIS0 spectra extracted from several inner regions of the PSF against those from an outer region in the outskirts of the PSF, where pileup is negligible. 
    (a) MCG~--6-30-15. 
    The inner region was taken from $r=0$--$60$ pixels. while the outer region was from $r=60$--$120$ pixels. 
    (b) GX339--4. The inner regions were taken from $r=0$--$25$, $25$--$60$, $60$--$75$, $75$--$90$, and $90$--$120$ pixels, while the outer region was from $120$--$240$ pixels.
 }
    \label{gx339}
\end{figure*}

\section{Observed Effects on Pileup}
\label{Effectsofpileup}

\subsection{Spectral Hardening and Grade Migration}

We first demonstrate two effects of pileup; 
i.e., spectral hardening and grade migration, using real data. 
We use a Seyfert galaxy MCG~--6-30-15 
and a Galactic black hole binary GX~339-4 as a representative case, respectively of little and significant pileup.

\subsubsection{Spectral hardening}

As explained in subsection 2.2,
the PSF is independent of the incident energy for the entire CCD field of view without pileup. 
As an example, we analyzed the spectra of MCG --6-30-15. 
Figure 2 shows ratios of XIS0 spectra extracted from several inner
regions of the PSF against those from an outer region in the outskirts
of the PSF, where pileup is negligible. 
Figure 2 (a) and (b) are for MCG~--6-30-15 and GX339--4, respectively.
The ratios are normalized to be 1.0 at 1 keV.

We confirmed in figure \ref{gx339} (a) that the spectral 
shapes between the $1'$ circle and 1--2$'$ annulus regions agree 
with each other within $\sim$~5\% in the flux in all energy bands. 
On the contrary,
when spectra are affected by pileup as shown in figure \ref{gx339} (b), 
the ratio between inner and outer regions becomes energy-dependent.

It is clear how significantly the spectra differ with
each other depending on the choice of the region.
The pileup effect on spectral distortion can be noticed
in the following two features:
1) moderate hardening over the whole energy range,
and 2) rapid hardening above $\sim$ 7 keV, probably due to sharp
decrease in the quantum efficiency of the CCD.
In addition to these two features, line-like features are noticed
in the spectral ratio around the instrumental edges (Si K at 1.8 keV
in the XIS and Au M at 2.3 keV in the XRT).
We consider this feature is caused by either 
the ratio between the spectrum with smeared-edge features owing to pileup and the one without it 
or the change in the charge transfer efficiency under
a high count rate.
The latter is not a direct outcome of pileup, 
because, when a count rate is high, a part of the events
works as sacrificial charges to improve the charge transfer efficiency
for others (Gendreau 1995).
This shifts the edge structure in the energy spectrum to higher
energies, which mimics a line-like feature when a spectral ratio
is calculated.

\begin{figure*}[htbp]
    \begin{center}
      \includegraphics[width=0.65\textwidth]{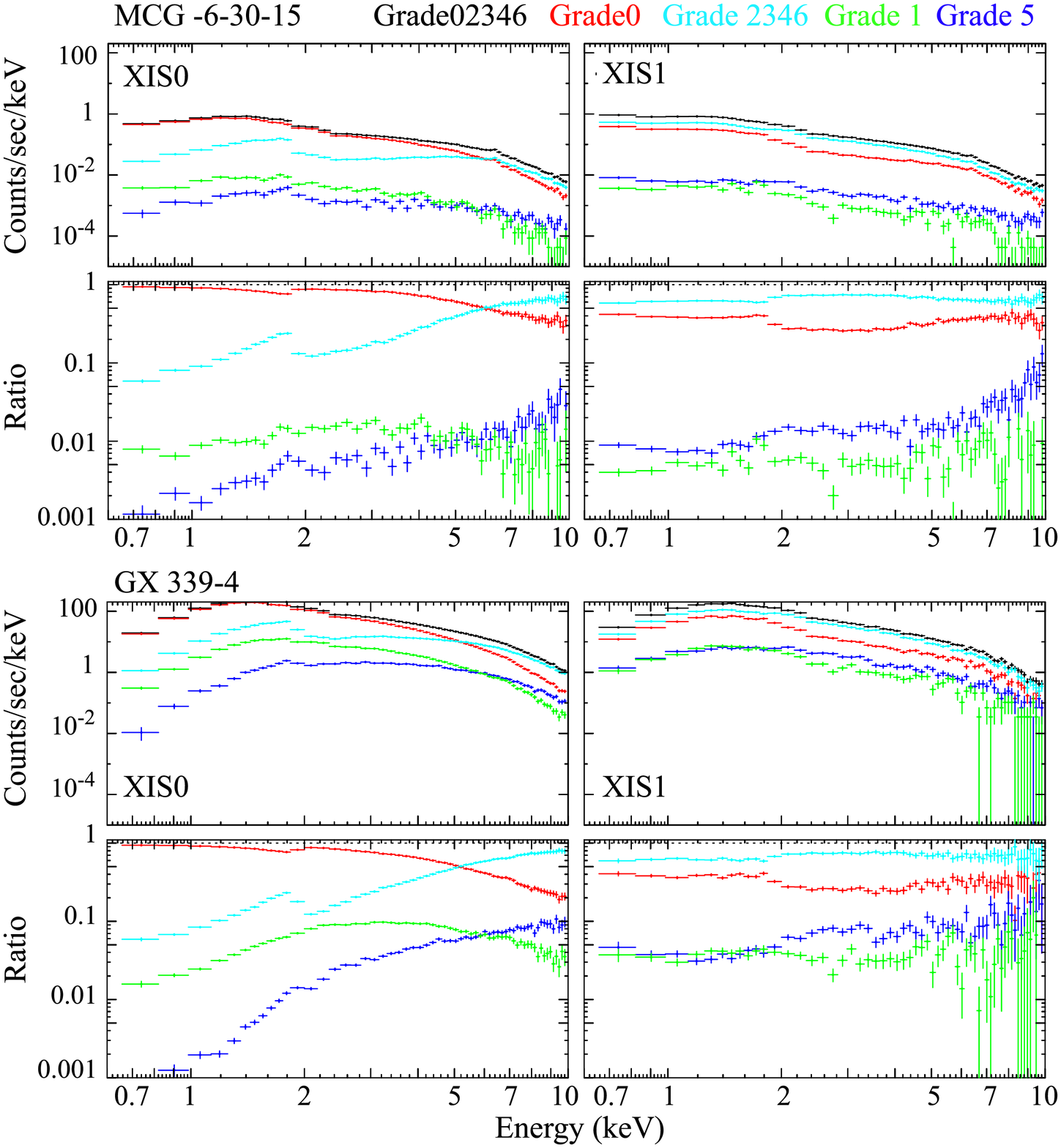}	
            \end{center}
   \caption{
   The detecter-response-convolved spectra and spectral ratio to the standard event (grade 02346) of MCG --6-30-15 (top) 
   and GX~339$-$4 (bottom). 
   The spectra of XIS0 and XIS1 are shown in left and right, respectively. 
   The black, red, cyan, green, and blue spectra refer to grade 02346, grade 0,
   grade 2346, grade 1, and grade 5 events, respectively. }
    \label{mcggx339}
\end{figure*}

\subsubsection{Grade migration}

To confirm the changes in grade branching ratio caused by pileup, 
we compared grade-sorted spectra and grade branching ratios of GX 339$-$4 and those of 
MCG~--6-30-15.
We extracted five spectra 
consisting of grade 02346, grade 0,  grade 2346, grade 1, and grade 5, 
which we present for XIS0 and XIS1 in left and right panels in figure \ref{mcggx339}, respectively,
for MCG~--6-30-15 (upper panels) and GX\,339--4 (lower panels).
To further clarify the energy dependence in grade branching ratio, 
we plot the spectral ratio 
against X-ray event grades (grade 02346).

In the MCG~--6-30-15 data, 
grade 0 events (red) and grade 2346 events (cyan) 
are dominant over the entire energy range, 
while the grade 1 (green) and grade 5 (blue) events are less than 1\% for both sensors. 
The ratios between grade 0 and grade 2346 of XIS0
are different between XIS0 and XIS1. 
In XIS0, 
a crossover is seen in the contribution of grade 0 and grade 2346 events; 
grade 0 is the largest below $\sim$ 6 keV and grade 2346 above $\sim$ 6 keV. 
In XIS1, on the other hand, 
no such crossover is seen and the grade 2346 events are the largest reaching at about 60--80\% in all energy bands. 
The difference is qualitatively understood by the fact that 
the distance 
between the point of photoelectric absorption in the depletion layer and the electrodes is longer for XIS1 than XIS0 for their opposite directions of illumination.

The apparent differences between MCG~--6-30-15 and GX 339$-$4 include 
(1) the crossover energy of grade 0 and grade 2346 in XIS0 
moves from 6 keV to 5 keV, 
(2) the ratio for grade 1 becomes larger in GX~339--4 ($\sim$5\%), than that in MCG~--6-30-15 ($\sim$1\%); 
the trend can be seen for both XIS0 and XIS1, and 
(3) the fraction of the grade 5 event is larger in GX~339$-$4 than MCG~--6-30-15, both for XIS0 and XIS1.

\subsection{Radial Profiles}

\begin{figure*}[htbp]
    \begin{center}
      \includegraphics[width=0.8\textwidth]{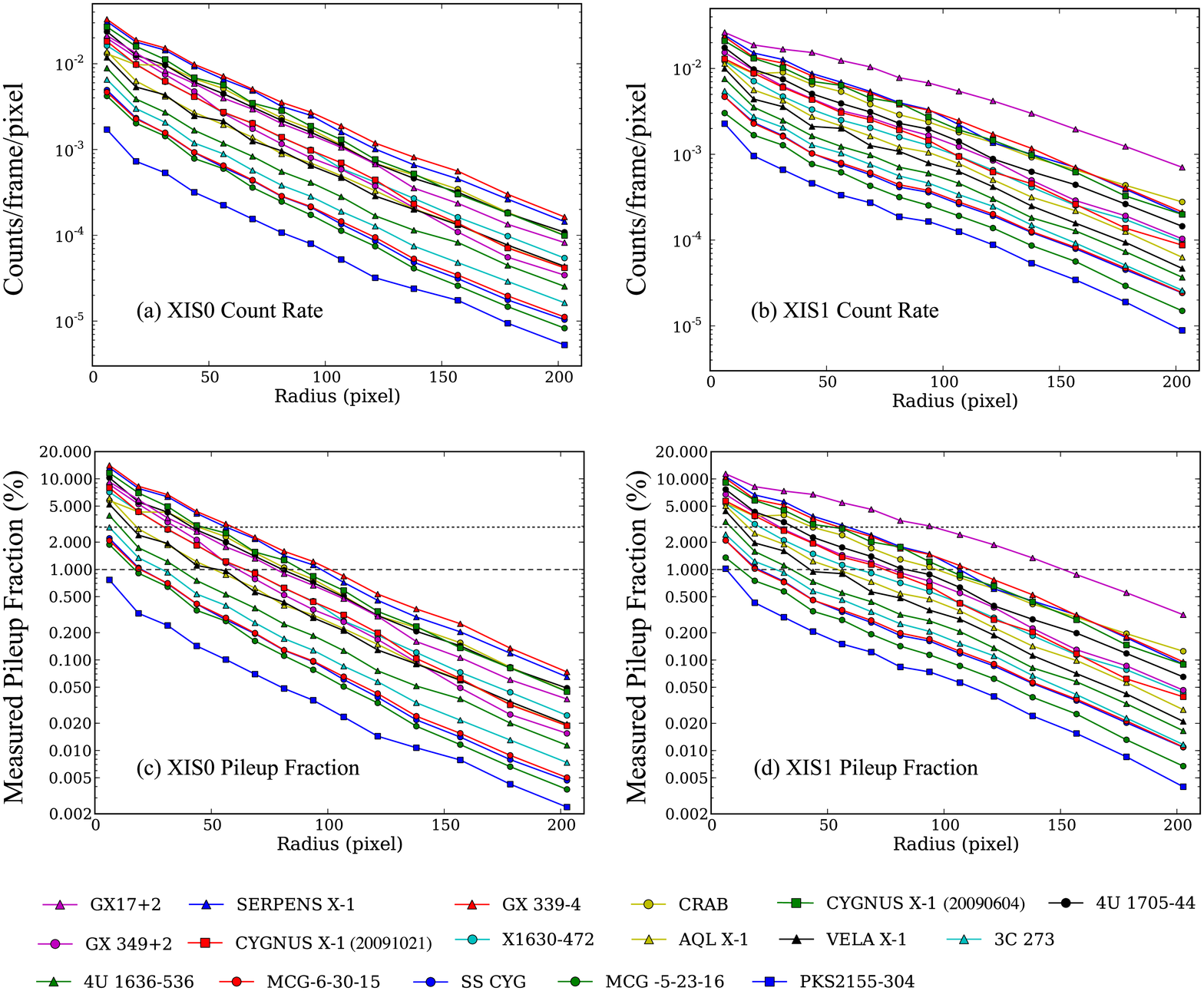}
            \end{center}
    \caption{(a) The radial profile of XIS0 of counts rate in a frame and a pixel for several sources; 
   the relation between target names listed in table 1 and the styles of the mark are shown below. 
   Their backgrounds are subtracted in a matter mentioned in section 3.2.2. 
   The profiles are averaged evenly-spaced in a linear scale in $r < 100$ pixel, 
and in a log scale in $r > 100$ pixel. 
   Although the error bars in Y-axis are actually plotted, they are smaller than the size of their marks. 
   (b) The same figure as (a) except for XIS1. (c) 
   The same figure as (a) except for a function of the measured pileup fraction. (d) The same figure as (b) except for a function of the measured pileup fraction. 
   The dotted lines and dashed lines refer to 3\% and 1\% of the measured pileup fraction, respectively. }
    \label{plfrac}
\end{figure*}

After applying the attitude correction (section 3.2.1), 
we extracted cleaned events (grade 02346) from the entire CCD regions. 
We use the energy ranges from 0.5 to 10.0 keV to maximize the photon statistics. 
Figure \ref{plfrac} shows radial profiles of surface brightness for XIS0 and XIS1
separately for each source. 

GX 17+2 has the largest counts per snap time in XIS1. 
It suffers a significant loss of counts at the PSF center, 
particularly in XIS1, which results in a broader profile than the others.

Multiplying the values of the vertical axis in figure \ref{plfrac}a and b by 9($=3\times3$), 
and then inserting them into $x$ in equation (1), 
we obtained $f^{obs}_{\rm{pl}}$, 
which is plotted in figure 4 (c) and (d) for XIS0 and XIS1. 
As long as the surface brightness is small, 
the pileup fraction is proportional to the surface brightness as expected from equation (1), 
Therefore, the profiles in figure \ref{plfrac} (a) and (b) are similar to those of (c) and (d), respectively.

\subsection{Pileup Effects Measured by Pileup Fraction}

The pileup fraction is the fundamental measure of pileup, 
while we cannot tell $f^{th}_{\rm{pl}}$ from $f^{obs}_{\rm{pl}}$ as mentioned in section 2.4. 
A more practical approach is 
to use the effects of pileup, 
such as spectral hardening and event migration as discussed in section 4.1, 
as a measure of pileup. 
Using the set of sources, 
we derive the quantitative relation between these effects as a function of pileup fraction.

\subsubsection{Spectral hardening}

As shown in figure 2, 
we have confirmed that the pileup causes the spectral hardening. 
To study the spectral hardening more systematically, 
we calculated hardness 
ratio as the count rate ratio between 7.0--10.0 and 0.5--10.0 keV 
at different radii for all sources (table 1), 
and plotted it against  $f^{\rm{obs}}_{\rm{pl}}$ in figure 5 (a) and (b) 
showing that our samples include sources with various intrinsic hardness.  
As expected, 
the hardness ratio increases monotonically as a function of increasing pileup fraction.

\begin{figure*}[htbp]
    \begin{center}
      \includegraphics[width=0.8\textwidth]{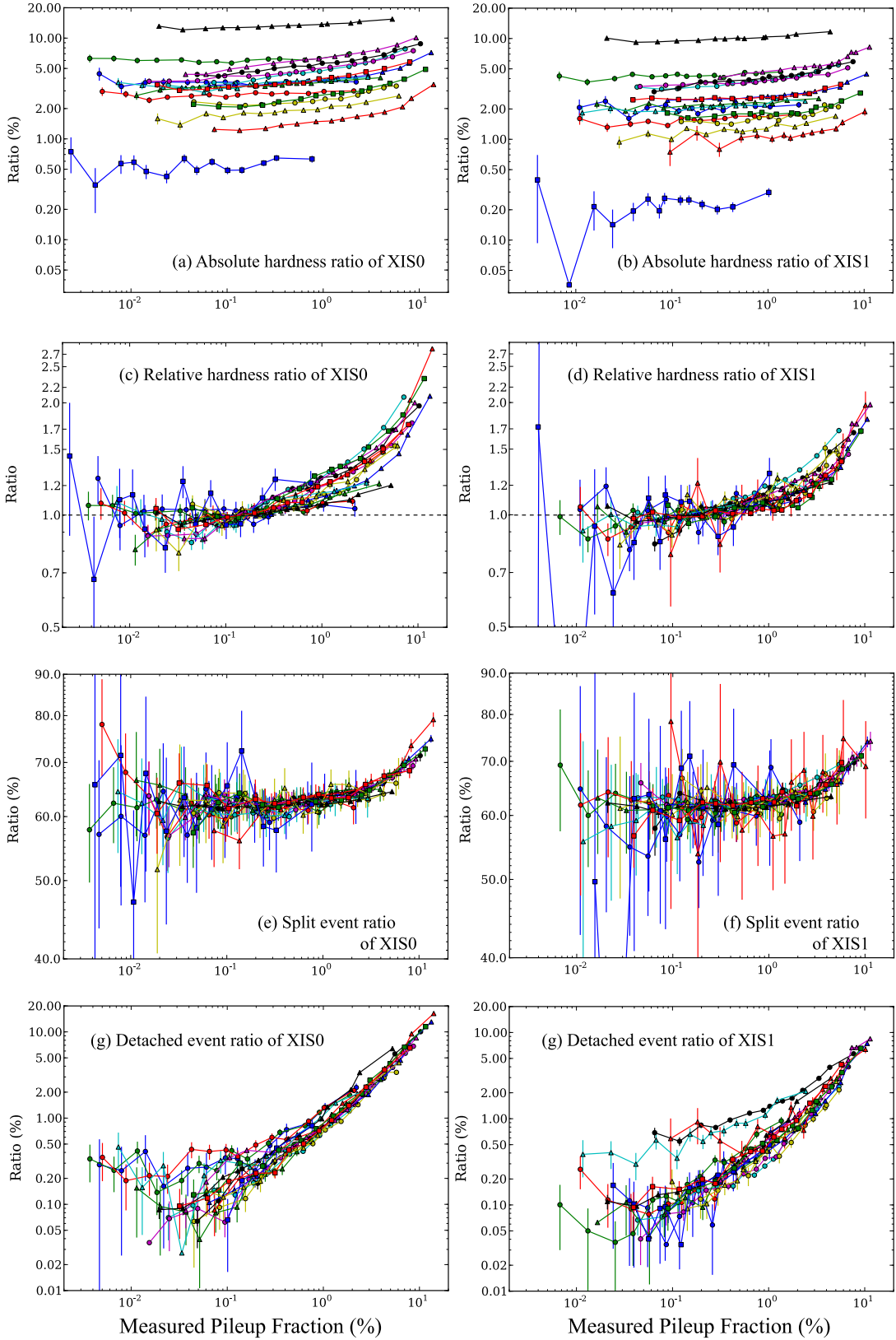}
            \end{center}
    \caption{(a) The profiles of the hardness ratios between 7.0--10.0 keV and 0.5--10.0 keV of the XIS0, 
    as a function of pileup fraction. 
    The symbols are the same as used in figure \ref{plfrac}. (b) the same as (a), except for XIS1.
    (c) the same as (a), except for being normalized to 1.0 at a pileup fraction of 0.1 \%. (d) the same as (c), except for XIS1. 
    (e) the ratio of split events (grade 2346) to the standard events (grade 02346), as a function of pileup fraction. 
    (f) the same as (e), except for the XIS1. 
    (g) the same as (e), except for detached events (grade 1). (h) the same as (g), except for the XIS1. }
    \label{alldata}
\end{figure*}

In order to compensate for the intrinsic spectral hardness of the sources, 
we normalized the hardness ratio at  $f^{\rm{obs}}_{\rm{pl}}$ of 0.1\%, 
which is shown in figure 5c and d. 
Despite the wide range of intrinsic spectral hardness, 
the relative spectral hardening shows the same trend. 
The hardening becomes significant
when $f^{\rm{obs}}_{\rm{pl}}$ reaches $\sim$ 1\%,   
and increases by a factor of $\sim$ 2 from $f^{\rm{obs}}_{\rm{pl}}$ of 0.1\% to 10\%. 
The degree of hardening is larger for XIS0 than XIS1 
because XIS0 has a larger effective area in the higher energy band.

\subsubsection{Grade migration --- split events} 

Since the grade branching ratio 
is sensitive to pileup, 
it can also be a good criterion on pileup, as shown in figure 3.  
In figure \ref{alldata} (e) and (f), 
we compare $f^{\rm{obs}}_{\rm{pl}}$ with the ``split event ratios'', the ratio of the split events (grade 2346)
to the standard events (grade 02346).  
The energy range from 7 to 10 keV is used 
because the split events are more imminent in the hard band (figure 3).

As shown in \ref{alldata}e and f, 
the split event ratios in 
all the 17 observations agree with each other within $\sim$ 10\%.
The branching ratios are fairly constant below $\sim$ 1\% of $f^{\rm{obs}}_{\rm{pl}}$. 
As $f^{\rm{obs}}_{\rm{pl}}$ becomes larger than $\sim$ 1\%, 
the split event ratios of XIS0 and XIS1 start to increase. 
Similarly to the changes in spectral hardness, 
the trend of the change of the split event ratio as a function of pileup fraction does not depend
on intrinsic hardness. 
The onset of the pileup effect can be seen at several \% of $f^{\rm{obs}}_{\rm{pl}}$.

\subsubsection{Grade migration --- detached events}

Finally, we show in figure \ref{alldata}g and h the ratio between the detached events (grade 1) 
and the standard events (grade 02346). 
Since, as shown in figure 3, grade 1 events are not significantly dependent on energy, 
we adopted the energy ranges of 3--10 keV. 

The profiles of XIS0 agree with each other within $\sim$ 10\%, 
which is reasonably understood by the idea that 
pileup causes the gradual increase of the detached events.  
Meanwhile, the profiles of XIS1 in 3C~273 and 4U~1705$-$44 deviate from the others by a factor of $\sim$ 5. 
We examined the images of the two observations, and found that 
it was caused by an asymmetric feature in the image of grade 1 events for some reasons. 
Aside from these outliers, 
the ratio of grade 1 events monotonically increases 
by orders of magnitude with the pileup fraction, 
in contrast to the profiles of 
relative hardness ratio and the split event ratio. 
The latter two appear to have a threshold in the pileup fraction to cause any changes (figure 5 c--f), whereas the detached event ratio appears to change from the lowest level of pileup (figure 5 g and h).
Thus, 
the fraction of the detached events is most sensitive to the pileup fraction.

\section{Practical case study}

\subsection{Pileup Fraction Criteria}

We have estimated the radius corresponding to 1\% and 3\% of $f^{\rm{obs}}_{\rm{pl}}$
(hereafter $R_{1\%}$ and $R_{3\%}$, respectively) 
by using the data points in figure 4 (c) and (d), 
and summarized them in table 2.
We also derived these values (in parentheses in table 2)
for the images before attitude correction. 
The estimate of the radii does not change so much, $\sim$ 1\% at most,  
whether the attitude correction discussed in section 3.2.1 was performed. 
The method is thus useful for images without the correction. 

We mask events within a certain radius, 
where the pileup effect is non-negligible. 
We use the radius for 1\% and 3\% pileup fraction 
($R_{1\%}$ and $R_{3\%}$, respectively) 
as representative cases of conservative and liberal choices. 
GX~339--4, for example, has $R_{3\%}$ and $R_{1\%}$ 
of 58.5 and 101.4 pixels in the XIS0, according to table 2. 
As shown in figure 2, 
the spectral ratio of 90--120 pixel ($\sim R_{1\%}$) to 120--240 pixel seems almost flat. 
Thus, $R_{1\%}$ can be a reasonable value for judging the onset of the pileup. 
In general, the appropriate mask radius should be chosen depending on science. 
At a larger $R$, 
the pileup effects decrease. 
However, attention should be paid for systematic uncertainties 
besides pileup effects. 
We see such an example in section 5.3.

\begin{table*}
 \caption{The values of $R_{1\%}$ and $R_{3\%}$ after and before the attitude correction.}
 \label{sample2}
 \begin{center}
  \begin{tabular}{lrrrrrrr}
  \hline\hline
\multicolumn{1}{c}{Name} & \multicolumn{2}{c}{XIS0} & \multicolumn{2}{c}{XIS1}  & \multicolumn{2}{c}{XIS 3}  \\[2mm] 
            & \multicolumn{1}{c}{$R_{3\%}$$^*$} & \multicolumn{1}{c}{$R_{1\%}$$^*$} & \multicolumn{1}{c}{$R_{3\%}$$^*$} & \multicolumn{1}{c}{$R_{1\%}$$^*$}  &\multicolumn{1}{c}{ $R_{3\%}$$^*$} & \multicolumn{1}{c}{$R_{1\%}$$^*$}  \\[2mm]                      
\hline
GX~17+2 & 39.0(38.9)$^\dagger$ & 78.3(78.0) & 94.3(94.4) & 151.2(151.1) & 42.3(42.1) & 81.7(80.9) \\
Serpens~X-1 & 55.5(55.4) & 97.9(97.9) & 58.5(57.7) & 106.0(106.6) & 54.8(54.8) & 99.6(99.6) \\
GX~339$-$4 & 58.4(58.5) & 101.6(101.4) & 53.6(53.7) & 110.8(111.0) & 76.4(76.5) & 117.3(117.2) \\
Crab & 43.2(43.4) & 83.4(83.3) & 42.0(42.0) & 97.0(97.1) & 39.7(39.9) & 82.0(81.8) \\
Cygnus~X-1(\#1) & 46.0(44.9) & 88.6(88.8) & 49.4(49.5) & 101.1(101.0) & 54.5(54.7) & 92.5(92.3) \\
4U~1705$-$44 & 41.2(41.1) & 80.2(80.4) & 35.3(35.2) & 84.9(83.5) & 40.5(40.2) & 80.3(79.9) \\
GX~349+2 & 34.3(34.2) & 61.2(61.3) & 28.7(28.6) & 77.4(77.4) & 36.6(36.5) & 63.4(63.5) \\
Cygnus~X-1(\#2) & 28.9(29.0) & 64.9(64.8) & 27.4(27.4) & 74.6(74.9) & 32.0(32.1) & 70.7(70.6) \\
4U~1630$-$472 & 28.7(28.7) & 65.1(65.0) & 20.5(20.4) & 63.5(63.4) & 30.4(30.2) & 69.2(69.4) \\
Aql~X-1 & 16.3(16.6) & 50.9(51.1) & 11.6(11.3) & 53.5(54.3) & 15.3(15.1) & 51.1(50.9) \\
Vela~X-1 & 8.4(6.4) & 53.0(53.1) & -(-) & 40.2(39.3) & 7.2(3.8) & 53.4(53.7) \\
3C~273 & 5.0(7.2) & 27.5(29.0) & 19.1(20.5) & 27.3(28.2) & 3.4(5.4) & 28.9(29.1) \\
4U~1636$-$536 & 2.8(3.5) & 37.2(36.8) & 4.0(4.3) & 34.8(34.6) & 2.3(1.9) & 36.1(36.2) \\
MCG~--6-30-15 & -(-) & 19.3(19.4) & -(-) & 19.1(19.5) & -(-) & 14.6(15.0) \\
SS~CYG & -(-) & 19.8(20.1) & -(-) & 19.7(20.6) & -(-) & 15.3(15.3) \\
MCG~--5-23-16 & -(-) & 15.4(15.3) & -(-) & 8.3(7.9) & -(-) & 18.3(18.4) \\
PKS~2155$-$304 & -(-) & -(-) & -(-) & 5.5(6.2) & -(-) & -(-) \\
\hline\hline
\end{tabular}
\end{center}
\begin{itemize}
\item[$^*$]  The radius at 3\% and 1\% pileup fraction, in an unit of pixel.
\item[$^\dagger$] The values in parenthesis are obtained without the attitude correction as mention in subsection 3.2.1.  
\end{itemize}
\end{table*}

\begin{figure*}[htbp]
    \begin{center}
      \includegraphics[width=0.78\textwidth]{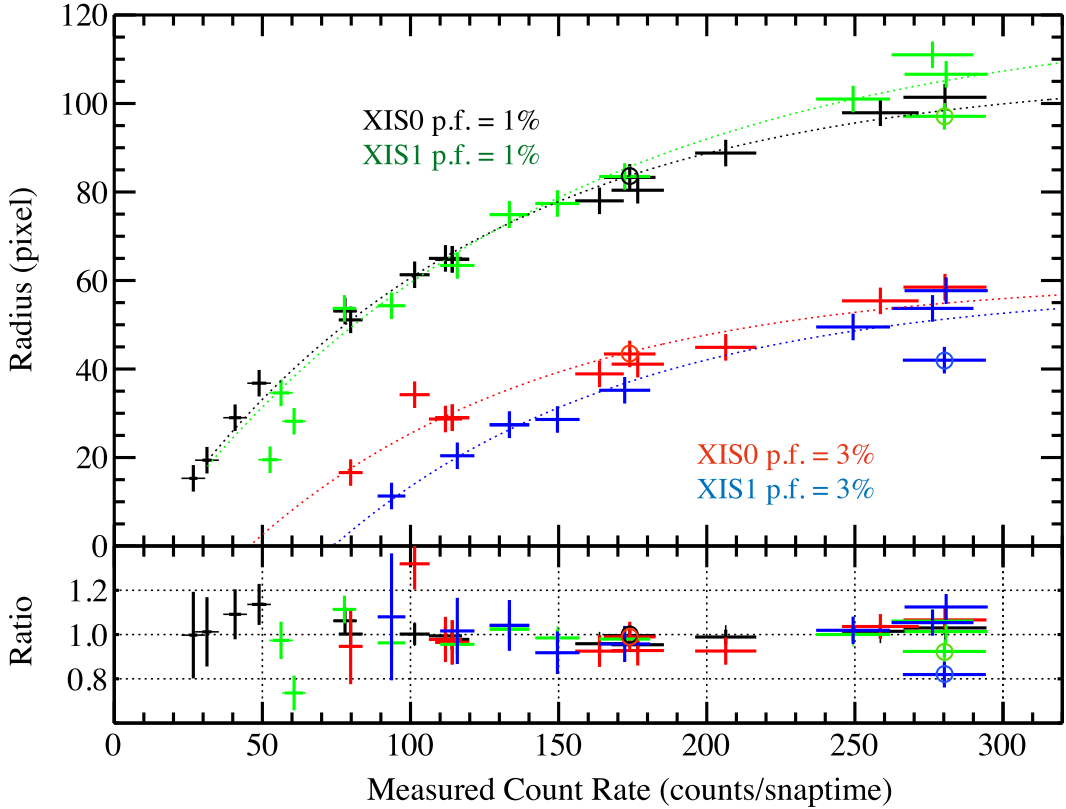}	
            \end{center}
    \caption{(top) The relation between measured counts per snap time (products of snap times by count rates as listed in table 1) over 0.5--10.0 keV and $R_{3\%}$ and $R_{1\%}$. 
    The black and red points correspond to $R_{1\%}$ and $R_{3\%}$ of XIS0, respectively; 
    whereas the green and blue ones correspond to those of XIS1, respectively. 
    The best-fit models are overlaid as dotted curves in the same color as the data.  
    The data of the Crab Nebula are marked as circle to distinguish them from the others. 
    (bottom) The ratios between the data to the best-fit models are shown.}
    \label{approxmodel}
\end{figure*}

Figure \ref{approxmodel} shows the relation 
between the measured counts per snap time with the values of $R_{3\%}$ and $R_{1\%}$ listed in table 2. 
The values of $R_{1\%}$ in XIS0 reaches almost zero at $\sim$ 10 counts per snap time.
To make it possible for general users 
to estimate the pileup extent before performing an observation, 
we quantified $f^{\rm{obs}}_{\rm{pl}}$ as a function of the measured counts per snap time, 
by fitting these values with the following phenomenological equation: 
\begin{equation}
\mathrm{Radius}\left( x \right) = a \left( 1 -  \exp \left( b \left( x - c \right) \right) \right) ~~  \mathrm{[pixel]},
\end{equation}
where $x$ is the measured counts per snap time, 
while $a$, $b$, and $c$ are free parameters. 
We took into account a systematic error of 10\% for the count rates, 
and $\pm$5 pixel as a systematic error for the radius, 
considering the intrinsic variation of the sources and individual spectral shapes. 
The data that show the radius less than 10 pixel and measured count rate larger than 320 counts snaptime$^{-1}$ 
are not used for fitting. 
In addition, only for the data with a pileup fraction of $3\%$,  
those that show a measured count rate smaller than 70 counts snaptime$^{-1}$ are not used. 
Since the observed PSF of the Crab Nebula is slightly wider due to its angular size of $\sim1'$, its values are not used for fitting. All the fits became acceptable,
and the resultant parameters are summarized in table 3.
We plotted the ratios to the best-fit model in the bottom of figure \ref{approxmodel}, 
showing an agreement to the data within $\sim$ 20\%.

A quick way to find suitable observation modes for a bright source, 
is to adjust window modes or burst options to 
realize that counts per snap time over 0.5--10 keV becomes below $\sim$ 10 counts per snap time.   
After observing it, 
then you can approximately know the extent of the pileup 
by comparing the measured counts per snap time with the radius in figure \ref{approxmodel}, or equation (2).

\begin{table}
 \caption{The fitting results of $R_{3\%}$ and $R_{1\%}$ as a function of measured surface density.}
 \label{fit}
 \begin{center}
  \begin{tabular}{crrrrr}
  \hline\hline
\multicolumn{1}{c}{} & \multicolumn{2}{c}{XIS0} & \multicolumn{2}{c}{XIS1}  \\[1mm] 
                                                                &  ($R_{3\%}$)$^*$ & ($R_{1\%}$)$^*$ & ($R_{3\%}$)$^*$ & ($R_{1\%}$)$^*$                                                    \\[2mm] 
\hline
                                 a$^\dagger$ &   $60.7\pm2.2$ & $107.3\pm3.8$  & $58.7\pm2.0$ & $121.0\pm4.9$     \\[1mm] 
                                 b$^\ddagger$ &  $-10$$^\|$   & $-9.2\pm0.7$     &  -10$^\|$ & $-7.5\pm0.7$                   \\[1mm] 
                                 c$^\S$  &  $45.7\pm5.9$ &  10$^\|$             & $74.9\pm5.3$ & $10$$^\|$                  \\[1.5mm] 
                  $\chi$(d.o.f) &  10.8(8) & 6.4(12)  &  14.3(7) & 12.9(8)                    \\[1mm] 
\hline\hline
\end{tabular}
\end{center}
\begin{itemize}
\item[$^*$]  The fitting results of $R_{3\%}$ and $R_{1\%}$, respectively. 
\item[$^\dagger$] In an unit of pixel. 
\item[$^\ddagger$] In an unit of $10^{-3}$ counts$^{-1}$ frame pixel. 
\item[$^\S$] In an unit of counts frame$^{-1}$pixel$^{-1}$. 
\item[$^\|$] The values are unconstrained and are fixed to the value at the lowest bound.
\end{itemize}
\end{table}

\subsection{Application to a severely piled-up source} 

\begin{figure}[htbp]
    \begin{center}
      \includegraphics[width=0.48\textwidth]{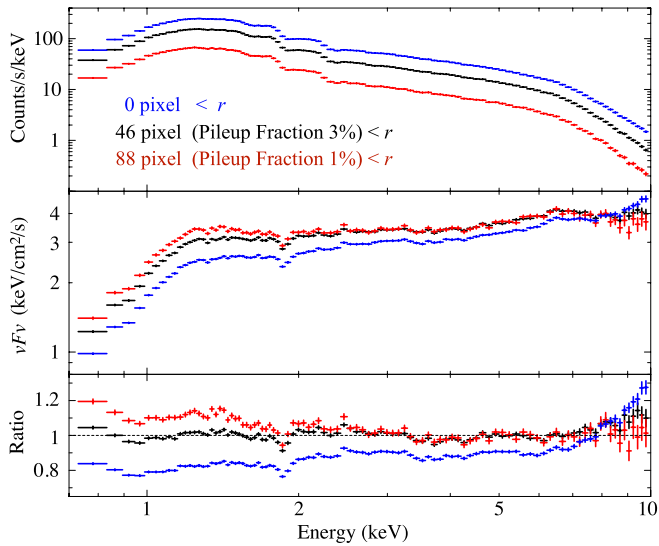}
            \end{center}
    \caption{(top) The detector-response-convolved spectra of the XIS0 of Cyg X-1 acquired on June 6 in 2009, accumulated from the region outside $R_{3\%}$ (46 pixel) and $R_{1\%}$ (88 pixel), presented in black and red, respectively. 
    As a references, the spectra taken from the whole region is also shown in blue. 
(middle) The same as top ones except for the detector-response-deconvolved spectra shown in a $\nu F \nu$ form. 
(bottom) The ratio to the model of {\tt diskbb + powerlaw + gaussian}, determined with the black spectrum.}
    \label{pl31check}
\end{figure}

To study uncertainties by using a region outside $R_{3\%}$ and $R_{1\%}$, 
we have compared spectra of the XIS0, taken from the observation of Cyg X-1 on June 2009 
in table 1, 
which is the most severely affected by pileup among the 25 observations of Cyg X-1 in the Low/Hard state
(for details see Yamada 2011).
As shown in table 2, 
the values of $R_{3\%}$ and $R_{1\%}$ in the XIS0 
are 46 and 88 pixels, respectively.  
Figure \ref{pl31check} shows the detector-response-convolved/deconvolved spectra accumulated 
from a whole CCD region, outside $R_{3\%}$ and outside $R_{1\%}$. 

The spectrum outside $R_{3\%}$ is 
fitted with a phenomenological model of {\tt diskbb + powerlaw + gaussian}, 
keeping the center energy of the gaussian fixed at 6.4 keV.
The resultant parameters are $T_{\rm{in}} = 0.22$ keV, 
normalization of {\tt diskbb} of 4.8 $\times 10^5$, 
$\Gamma = 2.0$, normalization of {\tt powerlaw} of 3.6 photons keV$^{-1}$ cm$^{-2}$ second$^{-1}$,  
and an equivalent width of 264 eV. 
The spectral ratio of the whole region (blue) to the model shows typical pileup features  
similar to the ratio in figure 2; i.e., 
the moderate hardening, the apparent decrease in flux, 
and the spectral steeping at the highest end around $\gtrsim $7 keV. 
On the other hand, the spectral ratio taken outside $R_{1\%}$ (red)
agrees with that outside $R_{3\%}$ (black) within $\sim$ 5\% above $\sim$ 2 keV. 
Thus, the spectrum outside $R_{3\%}$ 
is almost free from piled-up events.

\subsection{Other systematic uncertainties}

\begin{figure*}[b] 
    \begin{center} 
      \includegraphics[width=18cm]{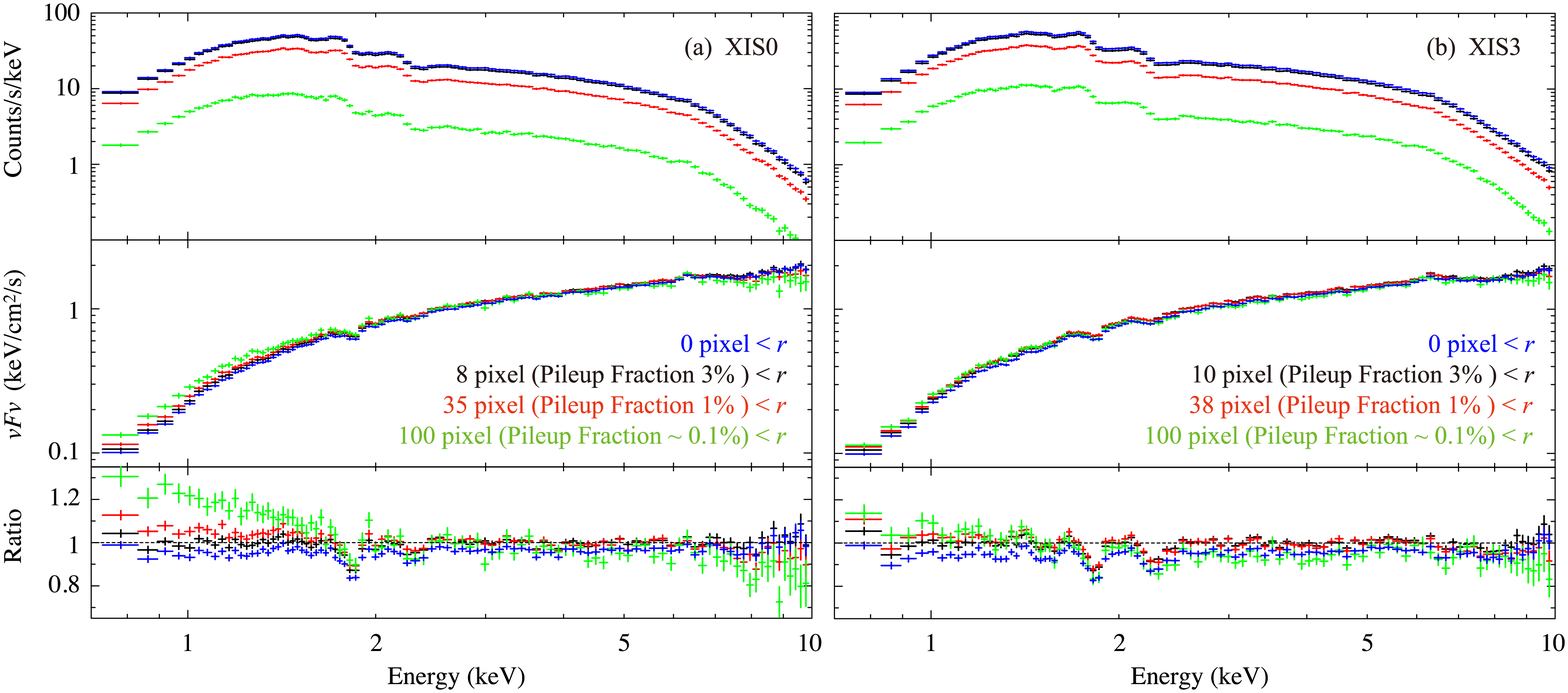}
    \caption{
    The same as figure \ref{pl31check},  
    but for the observation of Cyg X-1 taken in May 17th on 2007, 
    and for (a) XIS0 and (b) XIS3 data. 
    The spectra taken outside $R_{0.1\%}$ ($\sim$100 pixel)  are added in green. 
}
    \label{arfcheck}
    \end{center} 
\end{figure*}

For mitigating pileup effect, 
the larger mask radius is the better. 
However, at a large radius, 
we encounter different systematic uncertainties. 
In order to investigate these, 
we used the dimmest (OBSID=402072020) among the 25 Cyg X-1 observations
(see Yamada 2011 for details).
For the source, 
$R_{3\%}$ and $R_{1\%}$ of XIS0 and XIS1 
are 8 pixel and 35 pixel, and 10 pixel and 38 pixel, respectively. 
To illustrate systematic errors outside of $R_{1\%}$, 
we added $R_{0.1\%}$, which corresponds to $\sim$ 100 pixel, to as the mask radius and 
showed in figure \ref{arfcheck}. 
The spectra outside 8 pixel ($R_{3\%}$) are fitted with 
a model of {\tt diskbb + powerlaw + gaussian}, 
which is compared to the other spectra. 

Although all the spectra above $\sim$ 2 keV almost 
agree with each other within $\sim$ 5--10 \%, 
while the spectra taken outside of $R_{0.1\%}$ (green)
deviate by $\sim 20$\% around 0.7--2 keV only for the XIS0.  
Thus,  
it is likely that the deviation is 
caused by incomplete modeling of 
of the spatial distribution of the contamination accumulating on the CCD surface, 
because of thicker contamination in XIS0 than XIS3\footnote{See Figure 7. 17 of Suzaku technical description for AO7.\\ http://www.astro.isas.ac.jp/suzaku/proposal/ao7/\\suzaku\_td\_ao7.pdf}. 
This illustrates that 
a larger mask radius for a smaller pileup fraction is not always better in spectral analysis. 
We thus recommend to use either $R_{3\%}$ or $R_{1\%}$ as a practical solution. 

\section{Summary}

We have comprehensively studied pileup phenomena of the XIS,
by analyzing the actually observed data for several bright point sources. 
Using the surface brightness, 
the pileup fraction as the primary measure, 
we characterized grade-branching ratios of split events and detached events. 
We examined the effects on spectral analysis when excluding the central regions within $R_{3\%}$ and $R_{1\%}$.
Our results are summarized as follows: 

\begin{itemize}
\item Pileup causes spectral hardening in FI and BI CCDs,
as well as changes in the sharp spectral features such as atomic edges of instrumental response.
The changes in hardness is not dependent on intrinsic spectral hardness. 
\item Pileup increases the ratio of split events by $\sim$ 20\% 
from a pileup fraction of $\sim$ 1\% to $\sim$ 10\% in both FI and BI CCDs. 
On the other hand, the ratio of the detached events increases proportionally with the pileup fraction by $\sim$ several orders of magnitude from $\sim$ 0.1 and $\sim$ 10 \% of pileup fraction.
\item When surface brightness, spectral hardness, and split event ratio begin to increase, 
the pileup fraction starts to exceed $\sim$ 1\%,
which illustrate that the pileup fraction can be a good measure for estimating pileup. 
\item The relation between the measured surface brightness and the radius corresponding to pileup fraction of 3\% and 1\%, 
is useful to estimate the extent of pileup in a practical manner.   
\item Based on the observation of Cygnus X-1, the differences between the spectra accumulated outside $R_{3\%}$ and $R_{1\%}$ are less than 5\% in the 2--10 keV and $\sim$ 20 \%  in 0.5--2 keV. The latter might be caused by uncertainties of contamination modeling in the XIS. 
\end{itemize} 

Thus, we have phenomenologically compiled the pileup effects specific to the XIS + XRT systems and 
found a practical way of estimating pileup effects, 
which would be useful for those who analyze the bright sources taken with Suzaku, 
and even with similar CCD sensors to the XIS+XRT system, such as Soft X-ray Imager 
on board ASTRO-H (Tsunemi et al. 2010).  

\vspace{0.5cm}

The authors would like to express their thanks to Suzaku and the XIS team members, 
and anonymous referee for helpful comments and suggestions.  
The present work was supported by Grant-in-Aid for JSPS Fellows and RIKEN. 

{}


\begin{thebibliography}{}


\bibitem[Ballet  et al.(1999)]{Ballet1999}
Ballet, J., A\&A Suppl., 1999, 135, 371--381

\bibitem[Davis et al.(2001)]{Davis2001}
Davis, E. J., ApJ, 2001, 562, 575--582 

\bibitem[Ebisawa et al.(1996)]{Ebisawa1996} Ebisawa, K., Ueda, Y., 
Inoue, H., Tanaka, Y., \& White, N.~E.\ 1996, \apj, 467, 419 


\bibitem[Gendreau(1995)]{Gendreau1995} Gendreau, 
K.~C.\ 1995, Ph.D.~Thesis, Massachusetts Institute of Technology 


\bibitem[Kokubun et al.(2007)]{Kokubun2007}
Kokubun, M. et al. 2007, PASJ, 59, S53

\bibitem[Kotani et al.(2000)]{Kotani2000}
Kotani, T., Ebisawa, K., Dotani, T., Inoue, H., Nagase, F., Tanaka, Y., Ueda, Y., 2000, \apj, 539, 413--423 
[Erratum] \ 2006, \apj, 651, 615

\bibitem[Koyama et al.(2007)]{Koyama2007}
Koyama, K. et al. 2007, \pasj, 59, S23

\bibitem[Kubota et al.(2007)]{Kubota2007}
Kubota, A. et al. 2007, \pasj, 59, S185

\bibitem[Lumb  et al.(1991)]{Lumb1991}
Lumb, D. H., Berthiaume, G. D., Burrows, D. N., Garmire, G. P., Nousek, J. A.,
1991, Exptl. Astron., 2, 179--201

\bibitem[Makishima et al.(2008)]{2008PASJ...60..585M} 
Makishima, K., et al. 2008, \pasj, 60, 585 


\bibitem[Mitsuda et al.(2007)]{Mitsuda2007}
Mitsuda, K. et al. 2007, \pasj, 59, S1

\bibitem[Serlemitsos et al.(2007)]{Serlemitsos2007}
Serlemitsos, T.  et al. 2007, \pasj, 59, S9

\bibitem[Takahashi et al.(2007)]{Takahashi2007} Takahashi, T., Abe, 
K., Endo, M., et al.\ 2007, \pasj, 59, S35 

\bibitem[Tawa et al.(2008)]{Tawa2008}
Tawa N., et al. 2008, PASJ, 60, S11

\bibitem[Tsunemi et al. (2010)]{Tsunemi2010}
 Tsunemi, H., Hayashida, K., Tsuru, T.~G., et al.\ 2010, \procspie, 7732, 773210

\bibitem[Tsuneta et al.(1991)]{Tsuneta1991} Tsuneta, S., Acton, L., 
Bruner, M., et al.\ 1991, \solphys, 136, 37 


\bibitem[Uchiyama et al.(2008)]{Uchiyama2008} 
Uchiyama, Y., et al. 2008, \pasj, 60, S35

\bibitem[Ueda et al.(1997)]{Ueda1997} Ueda, Y., Inoue, H., 
Tanaka, Y., et al.\ 1997, All-Sky X-Ray Observations in the Next Decade, 
141 


\bibitem[Yamada et al.(2009)]{2009ApJ...707L.109Y}
Yamada, S., et al. 2009, \apjl, 707, L109 

\bibitem[Yamada et al.(2011)]{Yamada2011}
Yamada, S.,  2011, Ph.D thesis, Dept. of Physics, University of Tokyo  

\bibitem[Yamaguchi et al.(2006)]{Yamaguchi2006} Yamaguchi, H., et 
al.\ 2006, \procspie, 6266, 626642


\end{thebibliography}
\end{document}